\documentclass[12pt,preprint]{aastex}

\newcommand{\kms}{~km~s$^{-1}$} 
\newcommand{\teff}{$T_{\rm eff}$}

\newcommand{\loge}{$\log \epsilon$}

\slugcomment{}

\shorttitle{Extremely metal-poor stars from SDSS/SEGUE}
\shortauthors{Aoki et al.}

\begin{document}


%\title{Low-metallicity end of the metallicity distribution function of Galactic halo stars}

\title{High-Resolution Spectroscopy of Extremely Metal-Poor Stars from
  SDSS/SEGUE: II. Binary Fraction}


\author{Wako Aoki\altaffilmark{1}}

\affil{National Astronomical Observatory, Mitaka, Tokyo
181-8588, Japan}
\email{aoki.wako@nao.ac.jp}
\altaffiltext{1}{Department of Astronomical Science, School of Physical Sciences
, The Graduate University of Advanced Studies (SOKENDAI), 2-21-1 Osawa, Mitaka,
Tokyo 181-8588, Japan}

\author{Takuma Suda\altaffilmark{2}}
\affil{National Astronomical Observatory, Mitaka, Tokyo
181-8588, Japan}
\email{takuma.suda@nao.ac.jp}
\altaffiltext{2}{Present address: Research Center for the Early Universe,
University of Tokyo, 7-3-1 Hongo, Bunkyo-ku, Tokyo 113-0033, Japan}

\author{Timothy C. Beers}
\affil{Department of Physics and JINA-CEE: Joint Institute for Nuclear Astrophysics -- Center for the Evolution of the Elements, 225 Nieuwland Science Hall, Notre Dame, IN 46656, USA}
\email{tbeers@nd.edu}

\author{Satoshi Honda}
\affil{Center for Astronomy, University of Hyogo, 407-2, Nishigaichi, Sayo-cho, Sayo, Hyogo 679-5313, Japan}
\email{honda@nhao.jp}


%\altaffiltext{1}{Visiting Astronomer, Cerro Tololo Inter-American Observatory.
%CTIO is operated by AURA, Inc.\ under contract to the National Science
%\altaffiltext{2}{Society of Fellows, Harvard University.}
%\altaffiltext{3}{present address: Center for Astrophysics,
%    60 Garden Street, Cambridge, MA 02138}
%\altaffiltext{4}{Visiting Programmer, Space Telescope Science Institute}
%\altaffiltext{5}{Patron, Alonso's Bar and Grill}


\begin{abstract}

The fraction of binary systems in various stellar populations of the
Galaxy and the distribution of their orbital parameters are 
important but not well-determined factors in studies of star formation,
stellar evolution, and Galactic chemical evolution. While
observational studies have been carried out for a large sample of
nearby stars, including some metal-poor, Population II stars, almost
no constraints on the binary nature for extremely metal-poor (EMP;
[Fe/H] $< -3.0$) stars have yet been obtained. Here we investigate the
fraction of double-lined spectroscopic binaries and carbon-enhanced
metal-poor (CEMP) stars, many of which could have formed as pairs of
low-mass and intermediate-mass stars, to estimate the lower limit of
the fraction of binary systems having short periods. The estimate is
based on a sample of very metal-poor stars selected from the Sloan
Digital Sky Survey, and observed at high spectral resolution in a
previous study by Aoki et al.  That survey reported three double-lined
spectroscopic binaries and 11 CEMP stars, which we consider along with
a sample of EMP stars from the literature compiled in the SAGA
database. We have conducted measurements of the velocity components
for stacked absorption features of different spectral lines for each
double-lined spectroscopic binary. Our estimate indicates that the
fraction of binary stars having orbital periods shorter than 1000 days
is at least 10\%, and possibly as high as 20\%, if the majority of
CEMP stars are formed in such short-period binaries. This result
suggests that the period distribution of EMP binary systems is biased
toward short periods, unless the binary fraction of low-mass EMP stars
is significantly higher than that of other nearby stars.

\end{abstract}




%%  Otherwise, LaTeX will issue an herror. 

\section{Introduction}\label{sec:intro}

Understanding the nature of the first generation of stars born in the
Galaxy, at zero metallicity, is a key requirement for exploration of the
formation of structure, stellar evolution, nucleosynthesis, and chemical
evolution in the early Universe. While searches for metal-poor stars in
the Milky Way and its satellite galaxies have detected a large number of
stars with very low metallicity, no object with zero metallicity has yet
been found \citep{beers05,frebel13,keller14}. This might indicate that
only intermediate- to high-mass stars were formed from metal-free gas
clouds, which would have already exploded as supernovae or evolved to
white dwarfs. Indeed, recent numerical simulation of star formation from
gas clouds with no metals predict that the typical mass of first stars
to be several tens of solar masses or higher \citep[e.g., ][]{hirano14, susa14}. 

%massive star formation (several tens solar
%masses) are predicted by recent simulations including REFERENCES.

Some models of star formation from metal-free gas clouds, however,
predict the possible formation of low-mass stars \citep[e.g.,
][]{nakamura02}. Among such studies, the formation of low-mass stars as
multiple systems has also been investigated. For example, Machida (2008)
predicted that the binary frequency is higher in low-mass stars formed
from gas clouds with extremely low metallicity, because the probability
of fragmentation is larger at lower metallicity, even with expected
smaller rotation energy of such clouds. A high frequency of binaries
with short periods is also predicted.

The importance of obtaining observational estimates of the binary
fraction of stars is generally recognized. In addition to the studies of
stars in the Solar Neighborhood \citep[e.g., ][]{duquennoy91}, the
binary nature of metal-poor stars has been investigated by multi-epoch
spectroscopic approaches \citep[e.g., ][]{carney94, jorissen98,
goldberg02}, and by speckle interferometric observations of relatively
wide systems \citep[e.g., ][]{rastegaev10}.

However, there is almost no constraint on the binary frequency for
extremely metal-poor (EMP; [Fe/H] $< -3.0$) stars, which might be
compared with theoretical predictions for the low-mass stellar binary
formation at zero or extremely low metallicity. Although the number of
known EMP stars has been greatly increased by recent large spectroscopic
surveys, it will take some time to obtain a full estimate of their
binary fraction based on radial-velocity monitoring studies, as have
already been carried out for less metal-poor stars \citep{carney94,
latham02}.

In this paper we estimate a lower limit on the binary fraction for EMP
stars, based on the frequency of double-lined spectroscopic binaries
and carbon-enhanced metal-poor (CEMP) stars exhibiting over-abundances
of heavy neutron-capture elements (the CEMP-$s$ stars; Beers \&
Christlieb 2005). We investigate the sample of very metal-poor stars
reported by \citet{aoki13} (hereafter, Paper I), who obtained
high-resolution spectra for 137 stars selected from the Sloan Digital
Sky Survey (SDSS; York et al. 2000) and the stellar sub-survey SEGUE
\citep{yanny09}, providing a homogeneous sample of very metal-poor
stars with [Fe/H]$<-2.5$, in particular near the main-sequence
turnoff. We also consider samples of double-lined spectroscopic
binaries and CEMP-$s$ stars reported in literature, based on the SAGA
database compilation \citep{suda08}. Although the sample size of EMP
stars is still insufficient for deriving definitive conclusions, this
is the first attempt to provide a constraint on their binary
frequency and orbital periods.

%\section{Sample selection}\label{sec:sample}

\section{Double-Lined Spectroscopic Binaries}\label{sec:sb2}

The sample discussed in Paper I includes three double-lined
spectroscopic binaries. The velocity components of these systems are
analyzed in \S~\ref{sec:sdss} below. 

In order to estimate the fraction of binaries among EMP stars, we
investigate the detection probability of double-lined spectroscopic
binaries, as a function of the quality of our spectra, for stars close
to the main-sequence turnoff (\S~\ref{sec:sb2_vel} and
\ref{sec:sb2_lumi}). For this purpose, the sample is modeled using
stellar density distributions based on isochrones, assuming an initial mass
function and adopting an appropriate stellar age (\S~\ref{sec:stellarmodel}).

\subsection{Double-Lined Spectroscopic Binaries in the SDSS/SEGUE Subaru
Sample}\label{sec:sdss}

The three double-lined spectroscopic binaries found in Paper I are
listed in Table~\ref{tab:sdss}. Two spectra obtained at different epochs
are available for each object, including the spectrum of SDSS~0817+2641
studied by \citet{aoki08}. The quality of the data obtained by the
``snapshot'' spectroscopy technique (which employs shorter exposure
times than most high-resolution follow-up approaches) is only moderate
($S/N\sim 30$). In order to measure the velocity components of the
systems, we stacked absorption features of different spectral lines. The
spectral lines used in this analysis are \ion{Mg}{1} 5172.68~{\AA},
\ion{Mg}{1} 5183.60~{\AA}, \ion{Fe}{1} 4383.55~{\AA}, \ion{Ca}{1}
4226.73~{\AA}, and \ion{Ba}{2} 4554.03~{\AA}. The lines used for the
stacking of each spectrum are listed in the table. Note that the
\ion{Ba}{2} line was used only for the first-epoch spectrum of
SDSS~J0817+2641. The \ion{Ca}{1}, \ion{Fe}{1}, and \ion{Ba}{2} lines
were not used for the second-epoch spectrum of this object, because the
data quality at shorter wavelengths was not sufficient.

\begin{figure}
\epsscale{.6}
\plotone{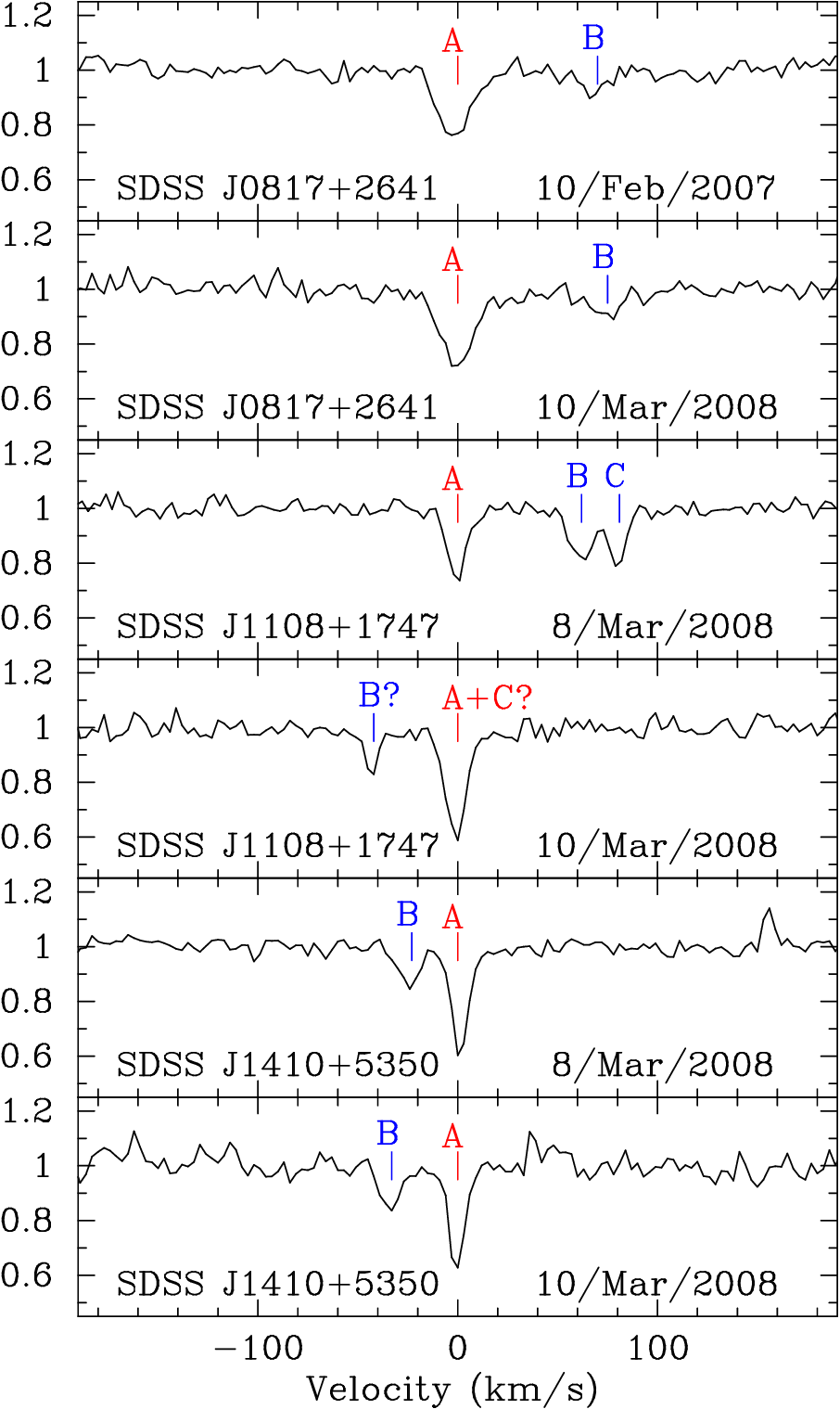}
\caption{Stacked spectra of absorption features for double-lined spectroscopic binaries in 
our sample.  The object name and
  observing date are presented in each panel. The absorption lines
  adopted are given in Table~\ref{tab:sdss}. The stronger and weaker
  absorption lines are labeled as components A and B,
  respectively. For the spectrum of SDSS~J1108+1747 obtained on March
  8, 2008, another component, C, is also identified. Only two
  components are identified in the spectrum of this object obtained on
  March 10, and the correspondence of the components is not certain.
\label{fig:sp}}
\end{figure}

Figure~\ref{fig:sp} shows the stacked spectra for individual exposures.
We identified two or three components in these spectra. The velocity
difference from the primary component, which is indicated as ``A'' in
the figure, were measured for each spectrum. The spectrum of
SDSS~1108+1747, obtained on March 8, 2008, exhibits three components,
indicating that this is (at least) a triple system. The other spectrum
of this star, obtained on March 10, 2008, exhibits only two components. A
comparison of line strengths found in the two spectra indicates that one
of the two velocity components, B or C, overlaps the primary (A) in the
March 10, 2008 spectrum. The heliocentric radial velocities of
individual components measured for the stacked spectra are given in
Table~\ref{tab:sdss}.

The properties of these stars are summarized in Table~\ref{tab:sb2}.
The effective temperature ({\teff}) and the iron abundances are taken
from Paper I. The {\teff} is approximately that of the primary star,
as discussed in Paper I. Assuming similar spectral features for the
two or three components in a given system, the ratios of the line
strengths (depths) in Figure~\ref{fig:sp} approximately indicate that
the luminosity ratios of the components are smaller than
4/1. According to theoretical isochrones of metal-poor stars
\citep[e.g., ][]{kim02}, this luminosity ratio corresponds to a pair
of main-sequence stars having a 400~K difference in {\teff}, or a
stellar pair comprising a subgiant and a main-sequence star having
almost the same effective temperature. 

\begin{figure}
\epsscale{.8}
\plotone{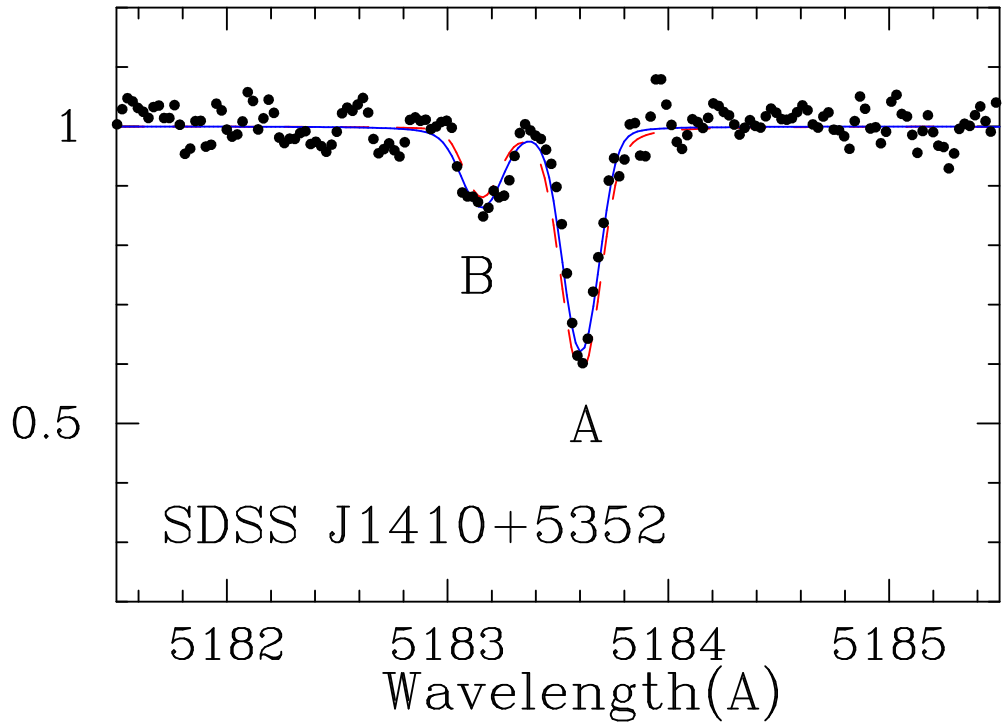}

\caption{Combined synthetic spectra of the \ion{Mg}{1} 5183.6~{\AA}
  line for the double-lined binary system SDSS~J1410+5350. The solid (blue) and
  dashed (red) lines indicate the cases of a main-sequence pair
  ({\teff}$=6300$ and 5800~K) and a subgiant (5900~K)--main-sequence
  (6500~K) pair. The flux ratio at this wavelength is assumed to be
  4/1, which is consistent with the estimate from the isochrone for
  the pairs. 
\label{fig:mg}}
\end{figure}

The above situation is demonstrated by \citet{aoki12} for the double-lined
spectroscopic binary G~166--45, which is a very metal-poor
([Fe/H]$=-2.5$) system of two main-sequence stars with 6300~K and
5900~K, with similar colors to SDSS~J1410+5350. Figure~2
of \citet{aoki12} shows that there are two cases of a main-sequence
stars' pair and a pair of main-sequence and subgiant stars that
satisfy the system's color as well as the mass ratio 
determined by long-term radial-velocity monitoring of the
system \citep{goldberg02}. Although the mass ratio of the binary
SDSS~J1410+5350 is still unknown, we estimate that the system
consists of two main-sequence stars with {\teff}$=6300$~K and
5800~K, or of a subgiant primary with 5900~K and a main-sequence
secondary with 6500~K. The combined synthetic spectra of the
\ion{Mg}{1} 5183.6~{\AA} line for the two cases of the system is
shown in Figure~\ref{fig:mg}, adopting the flux ratio of 4/1 as
estimated from the isochrone for very low metallicity
(Fig.~\ref{fig:iso}). Both cases account well for the observed spectrum
and are not distinguished by the current data quality. This
indicates that the above approach to estimate the components of the
binary system is basically appropriate, although the stellar
parameters of the two stars in the system are not fully
constrained by the fitting. In the case of a
subgiant--main-sequence pair, the mass ratio is about 0.9, while it is
closer to unity in the case of a pair of main-sequence stars.
The Mg abundance assumed in the calculation of the synthetic spectra
is {\loge}(Mg)$=4.9$ ([Mg/H]$=-2.7$). Adopting [Mg/Fe]$=+0.56$ estimated
for SDSS~J1410+5350 by \citet{aoki13}, this object is confirmed to
be extremely metal-poor ([Fe/H]$<-3$).

The other double-lined spectroscopic binaries listed in Table 2 are
taken from the literature, and are discussed in \S~\ref{sec:saga} and
\S~\ref{sec:disc} below.

\subsection{Detection Probability: Binary Separations}\label{sec:sb2_vel}

The first requirement for identifying an object as a double-lined
spectroscopic binary is that the velocity difference between the primary
and secondary stars, which reflects the separation of the two
components, is resolved in the spectra. The spectral resolution of our
data is $R=36,000$. Including the line broadening in the photosphere of
metal-poor, main-sequence turnoff stars, which is typically 5~{\kms},
the widths of observed lines are approximately 10~{\kms}. Hence, a
detectable velocity difference between the two components is 10~{\kms}
or larger. Indeed, the three binaries in our sample show line
separations larger than 20~{\kms}, as presented in Table~\ref{tab:sb2}.

Assuming a binary system with equal masses and zero eccentricity as the
simplest case, the two components would exhibit a radial velocity
variation with sine curves of the same amplitude. The fraction of the
orbital phase during which the velocity difference between the two
components is larger than 10~{\kms} is greater than 50~\% if the
amplitude (of each component) is larger than 7.5~{\kms}. The
corresponding orbital periods are about 1000~days or shorter. Hence, the
double-lined spectroscopic binaries detectable in our sample have rather
small separations (at most several AU). If a larger eccentricity is
assumed, binary systems having longer periods are detectable near their
periastron. However, the fraction of the orbital phase during which the
velocity difference becomes large is limited, and the probability to
detect such systems is small.

\subsection{Detection Probability: Luminosity Ratios}\label{sec:sb2_lumi}
\subsubsection{Detectable Luminosity Ratios}

Another requirement to detect a double-lined spectroscopic binary is
that the flux ratio between the two components is not so large that the
absorption lines in the spectrum of the secondary are completely swamped
by continuum light from the primary. In our spectra, with $S/N$ ratios
of 25--30, absorption lines with depths of 6--8\% are clearly
identifiable. Assuming the strengths of deep absorption features, such
as the \ion{Mg}{1} b lines (intrinsic absorption depth of $\sim 50$\%)
are the same between the two components, we can identify a double-lined
spectroscopic binary when the ratio of continuum flux between the
primary and secondary is 5/1 or smaller. Indeed, the continuum flux
ratios estimated for the three binaries in our sample are 4/1 or
smaller. Since the two components should have similar spectral types
when they are observed as a double-lined spectroscopic binary, the flux
ratio should be a good approximation of the luminosity ratio (see
\S~2.1).

In the following subsections, the probability to detect binary systems
based on stellar models that can explain the distribution of effective
temperatures of stars in our sample is explored.

\subsubsection{Stellar Models}\label{sec:stellarmodel}

Since the luminosity of a red giant is very sensitive to the mass for a
given stellar age, the probability to find a double-lined spectroscopic
binary for red giants is negligible. Hence, we here estimate the
probability to detect a double-lined spectroscopic binary for turnoff
stars as a function of {\teff}, adopting the stellar models of
\citet{kim02}.

\begin{figure}
\epsscale{.8}
\plotone{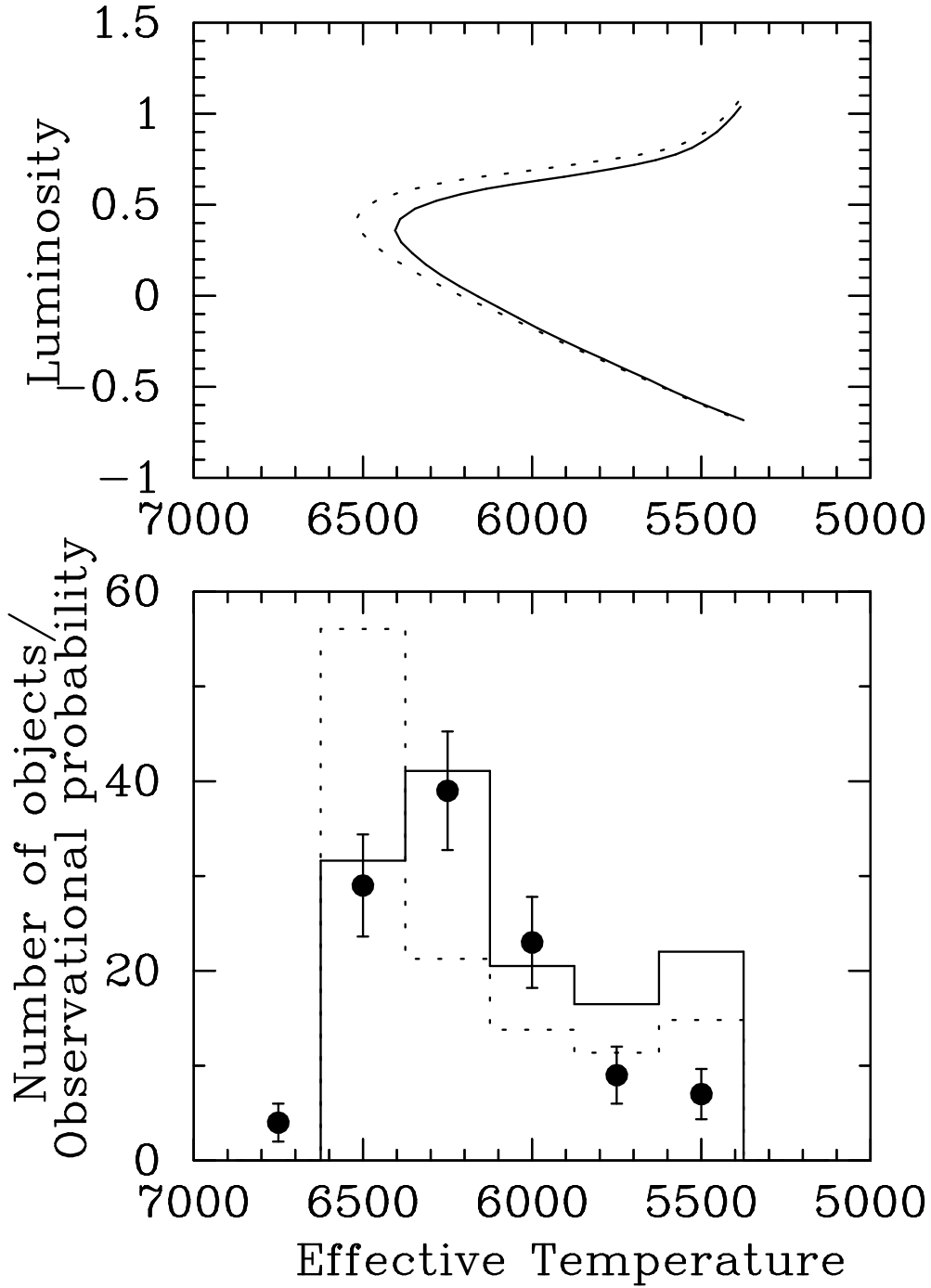}
\caption{Upper panel: The $Y^{2}$ isochrones for EMP stars
  ([Fe/H]$=-3.0$) with ages of 15.5~Gyrs (solid line) and 13.0~Gyrs (dotted
  line). Lower panel: The number of turnoff (main-sequence
  and subgiant) stars in our SDSS/SEGUE Subaru sample as a function of
  effective temperature (filled circles). The numbers expected for the
  above two isochrones are indicated by solid and dotted lines for ages
  of 15.5~Gyrs and 13.0~Gyrs, respectively.
\label{fig:iso}}
\end{figure}

For the first step, we estimate the probability to observe objects as a
function of evolutionary stages among turnoff stars.
Figure~\ref{fig:iso} (upper panel) shows the isochrones of
\citet{kim02} for [Fe/H]$=-3.0$, with enhanced $\alpha$-elements for
the ages of 13 and 15.5~Gyr. The probability to observe objects in a
flux-limited sample is proportional to $N \times L^{3/2}$, where $N$ and
$L$ indicate the stellar number density (calculated assuming a Salpeter
initial mass function) and stellar luminosity (here we ignore bolometric
corrections). 

The probability is calculated for each {\teff} bin of 250~K for
{\teff}$>5500$~K. Since main-sequence stars and subgiants are not
distinguished for our turnoff sample, we sum up the probabilities for
the two cases for each {\teff} bin. We note that the probabilities to
observe main-sequence stars and subgiants are comparable for
{\teff}$\geq$ 6000~K, while the probability to observe subgiants is
significantly higher than main-sequence stars for lower temperatures
(see below).

The lower panel of Figure~\ref{fig:iso} shows the results of the above
estimates, for two cases of stellar ages (the solid line: 15.5~Gyr;
the dotted line: 13~Gyr). As seen from their comparison, the
distribution of the highest {\teff} bin is sensitive to the adopted
age for the isochrone. The fraction of observed objects as a function
of {\teff} in our sample are shown by filled circles. We find that the
distribution for the 15.5~Gyr case agrees well with the observed data
for {\teff}$\geq 6000$~K. Hence, we here adopt the 15.5~Gyr isochrone
for the following estimates of the detection probability for
double-lined spectroscopic binaries. We note that, although this age
is higher than the currently adopted age of the universe ($\sim 13.8$
Gyrs; Ade et al. 2013), there could well be systematic offsets in the
calculations of stellar evolution and/or in the estimates of {\teff}
for our sample, which are not of consequence for the current
discussion. We also note that the number of stars with
{\teff}$<6000$~K in our sample is smaller than expected from the
number of objects with {\teff}$>6000$~K and the probability
distribution. Since, in our flux-limited sample, subgiants are
expected to dominate main-sequence stars in this temperature range,
this result indicates that subgiants with {\teff}$<6000$~K are
deficient in our sample compared to the prediction. This could,
however, also be due to the bias that our original sample of Paper I
includes a relatively small number of giants compared to turnoff
stars, which also partially affects the numbers of subgiants in the
sample.

\subsubsection{Probability to Detect Double-Lined Spectroscopic Binaries}

We now estimate the probability to detect double-lined spectroscopic
binaries as a function of {\teff} of the primary star, assuming a flat
(constant) mass distribution of the secondary. Under this assumption,
the probability is just the fraction of the mass range that satisfies
the secondary luminosity criterion (1/5th of the primary; see above). To
simplify the calculation, we here consider {\teff}$> 5375~K$ for both
the primary and secondary.

The probability estimated for each {\teff} bin (for the primary) is
given in Table~\ref{tab:prob}. The result is given separately for the
main-sequence and subgiant cases. The probability becomes lower with
increasing luminosity of the primary. Since the luminosity of a subgiant
is significantly higher than a main-sequence star, the mass range of the
secondary detectable as a double-lined spectroscopic binary is narrower.
This results in a lower probability for cooler subgiants. We also give
the fractions of main-sequence and subgiant stars expected to be
included in each {\teff} bin, which are proportional to $N \times
L^{3/2}$ (see above). The combined probability for main-sequence and
subgiant stars, weighted by the fractions of each, is given for each
{\teff} bin (see the table section labeled ``Total''). While the
probability to detect the secondary as a component of a double-lined
spectroscopic binary is higher for cooler main-sequence stars, the
fraction of main-sequence stars in a {\teff} bin is lower than that of
subgiants. As a result, the highest probability is found in the {\teff}
bins for 6000~K and 6250~K. Indeed, the three double-lined spectroscopic
binaries found in our sample have {\teff} in this range.

This probability is then multiplied by the number of objects in our
sample ($N_{\rm obs}$). The result is the expected fraction of
double-lined spectroscopic binaries detectable for each {\teff} bin,
if all of the primary stars are assumed to have secondaries in the
relevant temperature range.  The total number derived by this estimate
is 58\% of the observed stars. Hence, we can detect roughly half of
the binary stars as double-lined spectroscopic binaries, if the
components are subgiants or turnoff stars with {\teff}$\geq 5500$~K,
as well as having the required short period.

\subsection{The Fraction of Double-Lined Spectroscopic Binaries}

The fraction of detected double-lined spectroscopic binaries in our
turnoff sample is 2.8$\pm$1.6\% (three objects among 109 stars).
Although the sample size is still too small to derive any definitive
conclusion, we here attempt to give a constraint on the binary fraction
of EMP stars.

Given the above estimate for detection probability (\S~2.3), the
fraction of binaries in which the secondary is also a turnoff star is
$\sim $5--6\%. Taking into account the orbital phase for which the binary
system is detectable (\S \ref{sec:sb2_vel}), this fraction could be as
high as 10\%. This value indicates the fraction of binary systems that
have similar masses and orbital period of about 1000 days or shorter.

If a random distribution of the inclinations of binary systems, relative
to the line-of-sight ($\theta$), is assumed, a correction of the
detection probability is estimated to be $\int^{\pi/2}_{0}\cos\theta
d\theta/\int^{\pi/2}_{0} d\theta =2/\pi$. Hence, the correction to the
fraction of binary systems is the inverse of this ratio, about 3/2.

On the other hand, the fraction of binary systems in a flux-limited
sample could be enhanced by the effect that a binary system that is not
spatially resolved is brighter than a single star that has the same
brightness of the primary star \citep[\"{O}pik effect: ][]{trimble90}.
If we assume that the contribution of the secondary star to the flux
in the optical bands (e.g., the SDSS $g$-band) is typically 30~\% of the
primary, the volume covered by the flux-limited sample is $1.3^{3/2}\sim
1.5$. Hence, the correction to the fraction of binary systems is about
2/3, which is mostly compensated for by the correction for the effect of
inclination.

The estimate is dependent on the assumption of the eccentricity of
binary systems. However, the typical eccentricity ($e$) of binaries with
$P<1000$ days is 0.3 \citep{duquennoy91}, and the fraction of stars that
have $e>0.5$ is quite small (see also Goldberg et al. 2002). Hence, the
effect of the eccentricity assumed in the estimate should not have a
significant effect.

The estimate is also dependent on the assumption of the mass
distribution of secondaries. The secondary stars considered here are,
however, turnoff stars that have masses from 0.58 M$_{\odot}$ to 0.76
M$_{\odot}$. Hence, the result does not significantly change unless
there is a very sharp change of the mass ratio within such a small mass
range. Our calculation limits the lower mass of the secondary to be 0.58
M$_{\odot}$, because we here consider only turnoff stars. If lower-mass
stars are included as possible secondaries, the probability to detect
the secondary as a component of a double-lined spectroscopic binary in
our observation would become higher for cooler main-sequence cases.
However, as the fraction of such objects included in our sample is
small, the effect is also not expected to be large.

%uncertainties in this estiamte, excluding the uncertainty due to
%the smallness of the sample size, are

%\subsection{Binary fraction and orbital phases}

\section{Carbon-Enhanced Metal-Poor Stars}

The above estimate assumes that the currently observed object includes
two (or more) stars, the primary and secondary, that are still
shining.  However, when performing counts of the frequency of
low-mass star binaries, one should also account for systems in which we
are now observing the erstwhile secondary star, those that had
companions of higher mass that have already evolved to become faint
white dwarfs. In general, to estimate the fraction of such objects,
long-term monitoring for radial-velocity variations is
required. However, another useful probe for such systems is the
chemical peculiarity of stars, in particular the observed excesses of
carbon and neutron-capture elements, which would be expected to be
provided by mass transfer to the currently observed star from a
companion asymptotic giant branch (AGB) star in a relatively close
orbit. At very low metallicities, these stars are known as CEMP-$s$ stars
\citep{beers05}.  The initial mass of AGB stars that most efficiently
yield $s$-process elements are estimated to be 1--3~M$_{\odot}$
\citep{busso99}, although the operation of the $s$-process in EMP stars
is still under discussion \citep[e.g., ][]{suda04, lugaro12}.

As reported in Paper I, 9 turnoff stars and 2 red giants turned out to
be CEMP-$s$ stars in our total sample of 137 very metal-poor stars.
These objects are listed in Table~\ref{tab:cemps}. The fraction is about
8\%, both for turnoff stars and giants. This is a lower limit on the
fraction of CEMP-$s$ stars, because moderately carbon-enhanced (warm)
turnoff stars cannot be readily identified by our snapshot
spectroscopy.

Most of the CEMP-$s$ stars exhibit very large enhancements of carbon
([C/H] $ > -1$). In order to explain such large carbon excesses by mass
transfer from an AGB star across a binary system, the binary separation
would be expected to be relatively short ($\lesssim 10$~AU; see Fig. 8
of \citet{komiya07}). The lower limit on the separation of the binary
system required to form CEMP-$s$ stars is estimated to be
0.2--1~AU. If the separation was smaller, mass transfer by Roche-lobe
overflow would occur {\it before} the primary had evolved to an AGB star.

A recent study of the binary nature of CEMP stars \citep{starkenburg14}
confirms that almost all CEMP-$s$ stars can be regarded as binaries, with
average orbital periods 400--600~days. Hence, our sample of turnoff
stars includes binaries with periods $<3000$ days (possibly
$<1000$ days), in which the companion has already evolved to a white
dwarf with a fraction of 8\% (or higher).

%(HOWEVER, THIS DEPENDS ON THE ASSUMPTION OF THE SURFACE MIXING OF THE
%CEMP STAR AFTER MASS TRANSFER, SO I'M ACTUALLY NOT SURE FOR THIS
%ESTIMATE...)

\section{Binary Stars Reported in the Literature}\label{sec:saga}

We now inspect the frequencies of double-lined spectroscopic binaries
and CEMP-$s$ stars with extremely low metallicity, based on the sample
of EMP stars included in the SAGA database \citep{suda08}, which is a
compilation of stars with elemental-abundance data obtained from
high-resolution spectroscopy. The version of the SAGA database from
which we obtained the sample used for the present study contain data
published by 2012. We note that the sample of Paper I was not included
in the SAGA database.

In the SAGA database, we found 82 stars with [Fe/H] $ < -3.0$ and
{\teff}$>5500$~K, which represent main-sequence turnoff EMP
stars. Among them, four double-lined spectroscopic binaries have been
identified (Table~\ref{tab:sb2}). This fraction (4/82) is comparable
to, but somewhat higher than, that in our SDSS/SEGUE Subaru sample
(3/109). The reason for the higher fraction found for the SAGA sample
might be that high-resolution observations have been repeated for some
EMP stars included in the SAGA database, while the bulk of the sample
in Paper I is based on a single-epoch observation. That is, if the
spectrum of a SDSS/SEGUE Subaru star did not exhibit a clear signature
of it being a double-lined spectroscopic binary in the first-epoch
observation, it did not receive a second-epoch observation, resulting in
our possibly missing some of the bonafide double-lined binaries. If we
change the metallicity criterion from ]Fe/H] $< -3.0$ to [Fe/H] $<
-2.7$, the total number of stars increases to 156, but the number of
known double-lined spectroscopic binaries does not change. We also
suspect that double-lined binaries are sometimes excluded by studies of
the chemical abundances of EMP stars, due to the added complexity (and
uncertainty) of the analysis. Further complete assessment of the sample
of previously observed stars, including data not published, would be
quite useful for obtaining a more accurate estimate of the fraction of
double-lined spectroscopic binaries.

There are 149 stars with [Fe/H] $< -3$ for which [C/Fe] is determined in
the SAGA database. There are eight CEMP-$s$ stars with [Fe/H] $ <-3$
among them (Table~\ref{tab:cemps}). The fraction (8/149) is comparable
to that in our SDSS/SEGUE Subaru sample (11/137). We note that we adopt
a criterion of [Fe/H] $ < -3$ in the sample selection from the SAGA
database, while the sample of Paper I includes objects with slightly
higher metallicity, and indeed five stars among the 11 CEMP-$s$ stars
have [Fe/H] $ > -3$. The fraction of CEMP-$s$ stars among the entire
sample of CEMP stars is known to be lower in the metallicity regime
[Fe/H] $ < -3$, in which the CEMP-no\footnote{CEMP-no stars are defined
as CEMP stars without enhancements of neutron-capture elements; see
Beers \& Christlieb 2005. CEMP-no stars are not expected to be
associated with mass-transfer events; see \citet{norris13};
\citet{hansen14}; \citet{starkenburg14}, and Andersen et al. (in
preparation).} stars dominate \citep{aoki07}. The fractions of CEMP-$s$
stars for [Fe/H]$<-2.7$ and $<-2.5$ are 25/272 and 39/364, respectively,
which are as high as that found for our SDSS/SEGUE Subaru sample.

%Hence, the slightly lower fraction of CEMP-$s$ stars in
%the SAGA sample than that in our Paper I sample might be influenced by the
%small difference of the metallicity ranges covered by the two samples.
%
%HAVE YOU CONSIDERED RAISING THE LIMIT ON THE METALLICITY IN THE SAGA
%SEARCH TO MATCH OUR SAMPLE FROM SDSS/SEGUE ?  $=>$ I WILL TRY THE CASE
%WHEN WE SET THE CRITERION TO [Fe/H]=-2.7 AFTER UPDATE OF THE SAGA
%DATABASE THAT IS ONGOING.

\section{Discussion and Conclusions}\label{sec:disc}

We have detected three double-lined spectroscopic binaries among the 109
main-sequence turnoff stars studied in Paper I. The frequency of
low-mass star binary systems having comparable masses is 
estimated to be around 10~\%. Although the sample size is still small,
the result is supported by the fraction of double-lined
spectroscopic binaries among EMP turnoff stars in the SAGA database
(4/82). The orbital periods of these binary
systems would be shorter than about 1000 days. In addition, the
fraction of CEMP-$s$ stars, which would originally have had low-mass
(typically 1.5~M$_{\odot}$; Busso et al. 1999) star companions, is near
10~\% in both our Paper I sample and in the SAGA sample. Although the
estimate of binary separations (orbital periods) of such CEMP-$s$ stars
are dependent on the assumption of the mass transfer and surface mixing
of these stars, they could also be short ($P\lesssim 3000$ days), taking
into account their high [C/H] ratios (Table~\ref{tab:cemps}). Additional
measurements of orbital periods for these objects, based on
radial-velocity monitoring, is strongly desired.

The binary frequency of metal-rich stars in the Solar Neighborhood is
estimated to be about 50\% or higher, although that is not yet
considered well-determined. \citet{duquennoy91} reported that the
fraction of multiple stars in their sample (164 primary stars) is 43\%,
and the period distribution is unimodal with a median period of 180
years. According to their results, the fraction of binary systems with
period shorter than 1000 days is around 10\%. The fraction of such
short-period binaries among EMP stars estimated above is already as high
as 10\%, even though the searches for such binaries are clearly
incomplete. 

Concerning Population II stars, \citet{goldberg02} reported 34
double-lined spectroscopic binaries among the 1464 high proper motion
stars selected by \citet{carney94} (the Carney-Latham sample). A total
of 31 systems among these 34 have periods shorter than 1000~days, and the
remaining three have large eccentricity. The simple fraction is
34/1464 ($\sim 2.5$\%). About half of the double-lined spectroscopic
binaries of \citet{goldberg02} are, however, cool main-sequence stars
({\teff}$\lesssim 5250$~K). Such objects are not well-covered in
previous studies for very/extremely low-metallicity stars such as
studied in Paper I. On the other hand, we also need to take into
account the fraction of turnoff stars, having {\teff}$\gtrsim$5500~K,
included in the Carney-Latham sample, which is estimated to be about half,
according to Table 6 of \citet{carney94}. Hence, if the sample is
limited to turnoff stars, the fraction of double-lined spectroscopic
binaries is also about 2-3\% in the Carney-Latham sample.

%That is, about half of the
%whole sample as well as of the sample of \citet{goldberg02} is
%{\bf turnoff} stars.

%CAN WE SAY THAT DOUBLE-LINED SPECTROSCOPIC BINARIES ARE RICH IN emp
%STARS?

For the Carney-Latham sample, \citet{latham02} reported 181 single-lined
spectroscopic binaries based on their long-term radial velocity
monitoring survey, and concluded that the binary fraction among halo
stars was not lower than found for disk stars. The fraction of binaries
with $P<1000$ days is about 10\%. While binaries with long periods
($P>10000$ days) are not well-sampled by their study, their survey
should be close to complete for shorter periods. The fraction of short
period binaries in EMP stars is already as high as, or higher than, that
of the Carney-Latham sample, even though the searches for short period
binaries of EMP stars is still incomplete. This suggests that the high
fraction of short-period binaries could be a unique property found for
EMP stars.
 
%<-- NOT QUITE SURE I UNDERSTAND THIS ?  I THINK YOU NEED TO
%PUT IN THE EXPLICIT NUMBERS YOU ARE COMPARING TO HERE.

%  (THEIR MASS FRACTION... MANY
%BINARIES WITH LOW-MASS COMPANIONS?)
%
%Comparisons with nearby stars ([Fe/H]$\sim 0$) and  metal-poor stars ([Fe/H]$\sim -1$).
%
%Is our result compatible with the binary frequency/period distribution
%estimated by \citet{rastegaev10}, \citet{abt08}, etc.

One possible interpretation of the high fraction of short-period
binaries among EMP stars is simply that the binary fraction for EMP
stars is generally higher compared with metal-rich stars. This
hypothesis can be examined only by future investigations of wider binaries.
Another possibility is that the period distribution of binary systems
among EMP stars is biased toward shorter periods, as has been suggested
by \citet{rastegaev10} for metal-poor stars (but not for EMP stars). Further
long-term radial velocity monitoring or imaging studies with high
spatial resolution to distinguish binary components for EMP stars are
required to derive more definitive conclusions about the period
distributions of EMP binaries.

Numerical simulations of low-mass star formation from gas clouds with
various metallicities by \citet{machida08} demonstrate that the
fragmentation of clouds, which results in the formation of multiple
stellar systems, tends to occur in the clouds that have larger initial
rotational energy. The simulations indicate that the initial
rotational energy required for fragmentation depends on the
metallicity of the clouds. Fragmentation occurs in clouds with smaller
rotational energy at lower metallicity, because adiabatic cores form
at later phases of evolution in the clouds with lower metallicity. As
a result, binary (and multiple) systems with short periods are formed
with higher frequency at low metallicity, and the total fraction of
binary systems becomes higher. The high fraction of binary systems
with short periods among the EMP stars suggested by the present work
might well be accounted for by this effect. 

We note that, although a metallicity dependence of the
binary nature of stars has not been observed for more metal-rich 
populations \citep{latham02}, the metallicity range that has 
been previously inspected is [Fe/H]$\gtrsim -2$, higher than 
our EMP stellar sample. Further exploration of the properties of binary
systems in metal-poor stars, including the EMP population, is clearly
desired.

%\section{Summary and Concluding Remarks}

%   62.90257       107.0000      0.5878744    

\acknowledgments

This work is based on data collected at the Subaru Telescope, which is
operated by the National Astronomical Observatory of Japan. The data
obtained with Subaru/HDS are available on the Subaru Mitaka Okayama
Kiso Archive system (SMOKA: http://smoka.nao.ac.jp).  W.A. and
T.S. are supported by the JSPS Grant-in-Aid for Scientific Research
(S:23224004). S. H. is supported by the JSPS Grant-in-Aid for Young
Scientists (B:26400231). T.C.B.  acknowledges partial support from
grant PHY 08-22648: Physics Frontiers Center/Joint Institute for
Nuclear Astrophysics (JINA), and  PHY 14-30152; Physics Frontier Center/{}JINA
Center for the Evolution of the Elements (JINA-CEE), awarded by the U.S. National Science
Foundation.


%% in parentheses, after the keyword, but they will not be verified.

%{\it Facilities:} \facility{SDSS}, \facility{Subaru (HDS)}



\clearpage

\clearpage 

\begin{deluxetable}{lllcccl}
\tabletypesize{\scriptsize}
\tablewidth{0pt}
%\begin{center}
\tablecaption{Velocity Components of the SDSS/SEGUE Subaru Sample \label{tab:sdss}}
\tablehead{
\colhead{Object} & \colhead{Object name} &  \colhead{Obs. Date} & \multicolumn{3}{c}{$V_{\rm Helio}$ ({\kms})}  &  \colhead{Lines Used\tablenotemark{a} }
 \\
%\tablehead{
\cline{4-6} 
\colhead{} &  \colhead{} & &  \colhead{\#1} &  \colhead{\#2} &  \colhead{\#3} &  
}
%\tableline
\startdata
SDSS J0817+2641 & SDSS J081754.93+264103.8 & Feb. 10 2007   & $-$0.6 & $+$73.5  & \nodata & 1,2,3,5 \\
                &                          & March 10, 2008 & $+$2.7 & $+$72.0  & \nodata & 1,2 \\
SDSS J1108+1747 & SDSS J110821.68+174746.6 & March  8, 2008 & $-$88.7 & $-$25.4 & $-$9.27 & 1,2,3,4 \\
                &                          & March 10, 2008 & $-$28.4 & $-$70.6 & \nodata & 1,2,3,4 \\
SDSS J1410+5350 & SDSS J141001.77+535018.2 & March  8, 2008 & $-$138.6 & $-$164.5 & \nodata & 1,2,3,4 \\
                &                          & March 10, 2008 & $-$135.1 & $-$168.4 & \nodata & 1,2,3,4 \\
\enddata
\tablenotetext{a}{1: \ion{Mg}{1} 5172.68~{\AA}; 2: \ion{Mg}{1} 5183.60~{\AA}; 3: \ion{Fe}{1}  4383.55~{\AA}; 4: \ion{Ca}{1} 4226.73~{\AA}; 5: \ion{Ba}{2} 4554.03~{\AA}}
\end{deluxetable}

\begin{deluxetable}{llccccl}
\tabletypesize{\scriptsize}
\tablewidth{0pt}
\tablecaption{Double-Lined Spectroscopic Binaries\label{tab:sb2}}
\tablehead{
\colhead{Object} & \colhead{Object Name / Position (RA, Dec)} & \colhead{$V_{0}$} & 
\colhead{{\teff}(K)} & \colhead{[Fe/H]} & \colhead{$\Delta v$ ({\kms})} &
 \colhead{Remarks\tablenotemark{a}} 
}
%\begin{tabular}{llllccccc}
\startdata
\multicolumn{7}{c}{Objects in the SDSS/SEGUE Subaru Sample} \\
\hline
SDSS J0817+2641 & 08:17:54.93 ~ +26:41:03.8 & 16.012 & 6050 & $-2.85$ & 69--70 & 1,2 \\
SDSS J1108+1747 & 11:08:21.68 ~ +17:47:46.6 & 15.525 & 6050 & $-3.17$ & 42--79 & 1 \\
SDSS J1410+5350 & 14:10:01.77 ~ +53:50:18.2 & 16.015 & 6100 & $-3.42$ & 23--33 & 1 \\
\hline
\multicolumn{7}{c}{Objects from the Literature (SAGA Database)} \\
\hline
CS~22873--139   & 20:05:55.15 ~ --59:17:11.4 & 13.83 & 6300 & $-3.37$ & 53 & $P=19.166$~days; 3 \\       
CS~22876--032   & 00:07:37.46 ~ --35:31:16.7 & 12.84 & 6500 & $-3.66$ & 32 & $P=425$ days; 4,5 \\        
HE 1148$-$0037  & 11:51:14.98 ~ +00:54:11.1 & 13.61 & 5990 & $-3.51$ & 15:& 6,7\\       
HE 1353$-$2735  & 13:56:42.53 ~ --27:49:54.0 & 14.7  & 5900 & $-3.2$  & 30:& 8 \\       
\enddata
\tablenotetext{a}{References: 1: Paper I; 2: \citet{aoki08}; 3: \citet{spite00} ; 4:\citet{norris00}; 5: \citet{gonzalezhernandez08} ; 6: \citet{aoki09}; 7:\citet{sbordone10}; 8: \citet{depagne00}}
\end{deluxetable}

\begin{deluxetable}{llcccc}
\tabletypesize{\scriptsize}
\tablewidth{0pt}
%\begin{center}
\tablecaption{Probabilities to Detect Double-Lined Spectroscopic Binaries \label{tab:prob}}
\tablehead{
\colhead{{\teff} (K)} &  \colhead{5500} &  \colhead{5750} &  \colhead{6000} &  \colhead{6250} &  \colhead{6500} 
}
%\tableline
\startdata
\multicolumn{6}{c}{Main Sequence} \\
\tableline
probability\tablenotemark{a}   &  1.00 &  1.00 &  1.00 &  0.92 &  0.59 \\
fraction\tablenotemark{b}   &  0.13 &  0.35 &  0.52 &  0.55 &  0.47 \\
\tableline
 \multicolumn{6}{c}{Subgiant}\\
\tableline
probability\tablenotemark{a}   &  0.16 &  0.28 &  0.32 &  0.37 &  0.42 \\
fraction\tablenotemark{b}   &  0.87 &  0.65 &  0.48 &  0.45 &  0.53 \\
\tableline
 \multicolumn{6}{c}{Total}\\
\tableline
probability\tablenotemark{a}   & 0.27  &   0.53 &   0.68 &   0.67 &   0.50\\
$N_{\rm obs}$ &  7  &    9 &    23 &    39 &   29 \\
$N_{\rm pred}$ &  2 &   5 &    16 &    26 &   14 \\
%\tableline
%\end{tabular}
%\end{center}
\enddata
\tablenotetext{a}{The probability to detect a double-lined spectroscopic binary that has a turnoff star as secondary.}
\tablenotetext{b}{The fraction of main-sequence or subgiant stars among
the listed {\teff} bin.}
\end{deluxetable}

\begin{deluxetable}{llcccccl}
\tabletypesize{\scriptsize}
\tablewidth{0pt}
\tablecaption{CEMP-$s$ Stars\label{tab:cemps}}
\tablehead{
\colhead{Object} & \colhead{Object Name / Position (RA, Dec)} & \colhead{$V_{0}$} &
\colhead{{\teff}(K)} & \colhead{[Fe/H]} & [C/H] & [Ba/Fe] &
 \colhead{Remarks\tablenotemark{a}}
}
%\begin{tabular}{llllccccc}
\startdata
\multicolumn{8}{c}{CEMP-$s$ stars in the SDSS/SEGUE Subaru Sample} \\
\hline
SDSS J0002+2928   & SDSS J000219.87$+$292851.8 & & 6150 & $-3.26$ & $-0.63$ & 1.84 & 1\\
SDSS J0126+0607   & SDSS J012617.95$+$060724.8 & & 6900 & $-3.01$ & $+0.07$ & 3.20 & 1\\
SDSS J0711+6702   & SDSS J071105.43$+$670228.2 & & 5350 & $-2.91$ & $-0.97$ & 0.82 & 1\\
SDSS J0912+0216   & SDSS J091243.72$+$021623.7 & & 6150 & $-2.68$ & $-0.63$ & 1.30 & 1\\
SDSS J1036+1212   & SDSS J103649.93$+$121219.8 & & 5850 & $-3.47$ & $-1.63$ & 1.35 & 1\\
SDSS J1245$-$0738 & SDSS J124502.68$-$073847.1 & & 6100 & $-3.16$ & $-0.63$ & 2.09 & 1\\ 
SDSS J1349$-$0229 & SDSS J134913.54$-$022942.8 & & 6200 & $-3.24$ & $-0.23$ & 2.25 & 1\\
SDSS J1626+1458   & SDSS J162603.61$+$145844.3 & & 6400 & $-2.99$ & $-0.13$ & 1.69 & 1\\
SDSS J1646+2824   & SDSS J164610.19$+$282422.2 & & 6100 & $-3.05$ & $-0.53$ & 1.78 & 1\\
SDSS J1734+4316   & SDSS J173417.89$+$431606.5 & & 5200 & $-2.51$ & $-0.73$ & 1.61 & 1\\
SDSS J1836+6317   & SDSS J183601.71$+$631727.4 & & 5350 & $-2.85$ & $-0.83$ & 2.37 & 1\\
 \hline
\multicolumn{8}{c}{CEMP-$s$ Stars with [Fe/H]$<-3$ from the Literature
(SAGA Database)} \\
\hline
CS~22183--015   & 01:00:53.03 ~ $-$02:28:20.0  & & 5178 & $-3.17$ & $-1.07$ & 1.77 & 2 \\  
CS~22960--053   & 22:16:17.08 ~ $-$43:54:19.7  & & 5200 & $-3.14$ & $-1.09$ & 0.86 & 3 \\  
%CS~29493--090   &                            & & 4692 & $-3.13$ & $-2.55$ & 0.52 & 4 \\  
CS~29528--041   & 02:29:25.1  ~ $-$18:13:28.0  & & 6150 & $-3.30$ & $-1.85$ & 0.97 & 4 \\  
HE~1005--1439   & 10:07:52.3  ~ $-$14:54:21.0  & & 5000 & $-3.17$ & $-0.69$ & 1.06 & 3 \\
HE~1410--0004   & 14:13:04.63 ~ $-$00:18:32.5  & & 5605 & $-3.02$ & $-1.03$ & 1.06 & 5 \\  
SDSS~J0126+06   & SDSS J012617.95$+$060724.8    & & 6600 & $-3.11$ & $-0.19$ & 2.75 & 6 \\  
SDSS~J1036+1212 & SDSS J103649.93$+$121219.8    & & 6000 & $-3.2$  & $-1.73$ & 1.17 & 7 \\  
SDSS~J1349-0229 & SDSS J134913.54$-$022942.8    & & 6200 & $-3.0$  & $-0.18$ & 2.17 & 7 \\  
\enddata
\tablenotetext{a}{References: 1: Paper I; 2: \citet{lai08}; 3:
\cite{aoki07}; 4: \citet{sivarani06}; 5: \citet{cohen06}; 6: \citet{aoki08}; 7: \citet{behara10}}
\end{deluxetable}

\end{document}